\documentclass[conference]{IEEEtran}
\IEEEoverridecommandlockouts
\usepackage{amsmath,amssymb,amsfonts}
\usepackage{graphicx}
\usepackage{textcomp}
\usepackage{xcolor}
\usepackage{booktabs} 
\usepackage[linesnumbered,ruled,vlined]{algorithm2e}
\usepackage{booktabs}
\usepackage{amsmath}
\usepackage{graphicx}
\usepackage{subfigure}
\usepackage{cite}
\usepackage{multirow}
\usepackage{amsfonts,bm}
\usepackage{color}
\usepackage{braket}
\usepackage{mathtools}
\usepackage{enumerate}
\usepackage{enumitem}
\usepackage{empheq}
\usepackage{calc}
\usepackage{balance}
\usepackage{dsfont}
\usepackage{textcomp} 
\usepackage{color, colortbl}
\usepackage[first=0,last=9]{lcg}
\usepackage[colorlinks,linkcolor=black,citecolor=green,  urlcolor=black]{hyperref}
 
\def\BibTeX{{\rm B\kern-.05em{\sc i\kern-.025em b}\kern-.08em
    T\kern-.1667em\lower.7ex\hbox{E}\kern-.125emX}}
\begin{document}

\title{Reconstructing High-resolution Turbulent Flows Using Physics-Guided Neural Networks \\
}
\author{Shengyu Chen$^1$, Shervin Sammak$^2$, Peyman Givi$^2$, Joseph P. Yurko$^1$, Xiaowei Jia$^1$\\
$^1$Department of Computer Science, University of Pittsburgh\\
$^2$Department of Mechanical Engineering and Materials Science, University of Pittsburgh\\
\{shc160, shervin.sammak, peg10, jyurko, xiaowei\}@pitt.edu
}


\maketitle

\begin{abstract}
  Direct numerical simulation (DNS) of turbulent flows is computationally expensive and cannot be applied to flows with large Reynolds numbers.  Large eddy simulation (LES) is an alternative that is computationally less demanding, but is unable to capture all of the scales of turbulent transport accurately. Our goal in this work is to build a new data-driven methodology based on super-resolution techniques to reconstruct DNS data from LES predictions. We leverage the underlying physical relationships to regularize the relationships amongst different physical variables. We also introduce a hierarchical generative process and a reverse degradation process to fully explore the correspondence between DNS and LES data. We demonstrate the effectiveness of our method through a single-snapshot experiment and a cross-time experiment. The results confirm that our method can better reconstruct high-resolution DNS data over space and over time in terms of pixel-wise reconstruction error and structural similarity. Visual comparisons show that our method performs much better in capturing fine-level flow dynamics.

\end{abstract}


\section{Introduction}
 
Computational fluid dynamics (CFD) has proven to be a very effective research tool in a very wide variety of disciplines, including engineering, science, medicine and more \cite{CFD1995}. For its applications in  turbulent flows, however, the range of the temporal \&  spatial scales is too broad to be captured by brute force direct numerical simulations (DNS) \cite{Davidson2015}. Large eddy simulation (LES) provides an alternative, by filtering the small-scale scales of transport and concentrating on the larger scale energy containing eddies \cite{Sagaut05}. By this filtering, LES can be conducted on coarser grids as compared to those required by DNS. The penalty, understandably, is that LES generated data are of lower accuracy  compared to DNS. Appraisal of LES predictions and assessments of its fidelity as compared to DNS, have been of interest in the turbulence research community for the past several decades \cite{givi1989model,pope2001turbulent}. The objective of the present work is to build a new data-driven methodology to reconstruct DNS from LES data, which facilitates a more robust means of LES appraisal.

Machine learning, including super-resolution methods~\cite{Cheo2003SR}, have shown great success in reconstructing high-resolution data in a variety of commercial applications. For example, convolutional neural networks (CNNs) and their extensions, e.g., SRCNN~\cite{dong2014learning}, RCAN~\cite{zhang2018image}, and SRGAN~\cite{ledig2017photo}, have proven very effective in directly mapping low-resolution images to high-resolution images. The effectiveness of these methods mainly come from the power of CNNs in automatically extracting representative spatial features through deep layers. An alternative solution is to consider super-resolution as an inverse modeling problem~\cite{geiss2020invertible,mccann2017convolutional} with the constraints that the down-sampled version of the underlying high-resolution data should be consistent to the observed low-resolution data.

\begin{figure}[t]
\centering  
\subfigure[LES]{
\label{Fig.sub.1}
\includegraphics[width=4.1cm,height = 4.1cm]{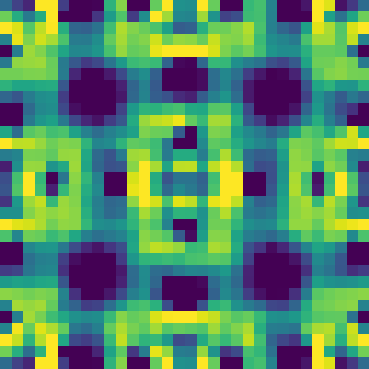}}\subfigure[DNS]{
\label{Fig.sub.2}
\includegraphics[width=4.1cm,height = 4.1cm]{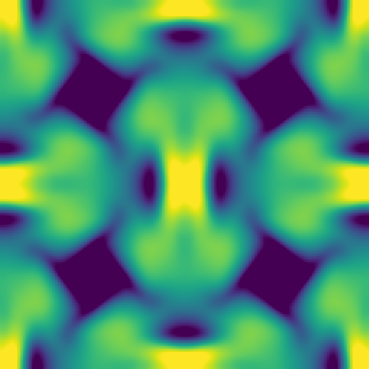}}
\caption{An example slice of Large Eddy Simulation (LES) and its corresponding Direct Numerical Simulation (DNS). This example is used to show the 
the difference between LES and DNS at certain spatial locations and certain time steps. }
\label{Sample}
\end{figure}

Super-resolution techniques are starting to be used in turbulence research \cite{liu2020deep,xie2018tempogan,fukami2019super}. However, there are several major challenges that must be overcome, before they can be employed for routine applications. 
First, turbulent flow data often exhibit significant variability. In the absence of underlying physical processes, machine learning models are prone to learning spurious patterns that fit statistical characteristics of available training data collected from a specific time period, but cannot generalize to other time intervals. This can be further exacerbated by limited training data. 
Second, existing super-resolution algorithms could have degraded performance in CFD because of the huge information loss caused by the large resolution gap. For example, LES data can be of more than 8 times lower resolution compared to DNS data along each axis. Hence, standard statistical interpolation methods may fail to capture fine-level flow dynamics resulting from underlying physical relationships and constraints. Third, existing machine learning  models are not designed to deal with the discrepancy between 
different simulation strategies (i.e. LES, DNS and/or others). In general, the available simulations at a coarser resolution are not simply a down-sampled version of high-resolution simulations. Consider the examples in Fig.~\ref{Sample}. It is obvious that the LES predictions on the coarser grids do not capture the flow patterns as compared to high-resolution DNS.

\begin{figure*}[htbp]
\begin{center}
\includegraphics[width = 1.03\linewidth]{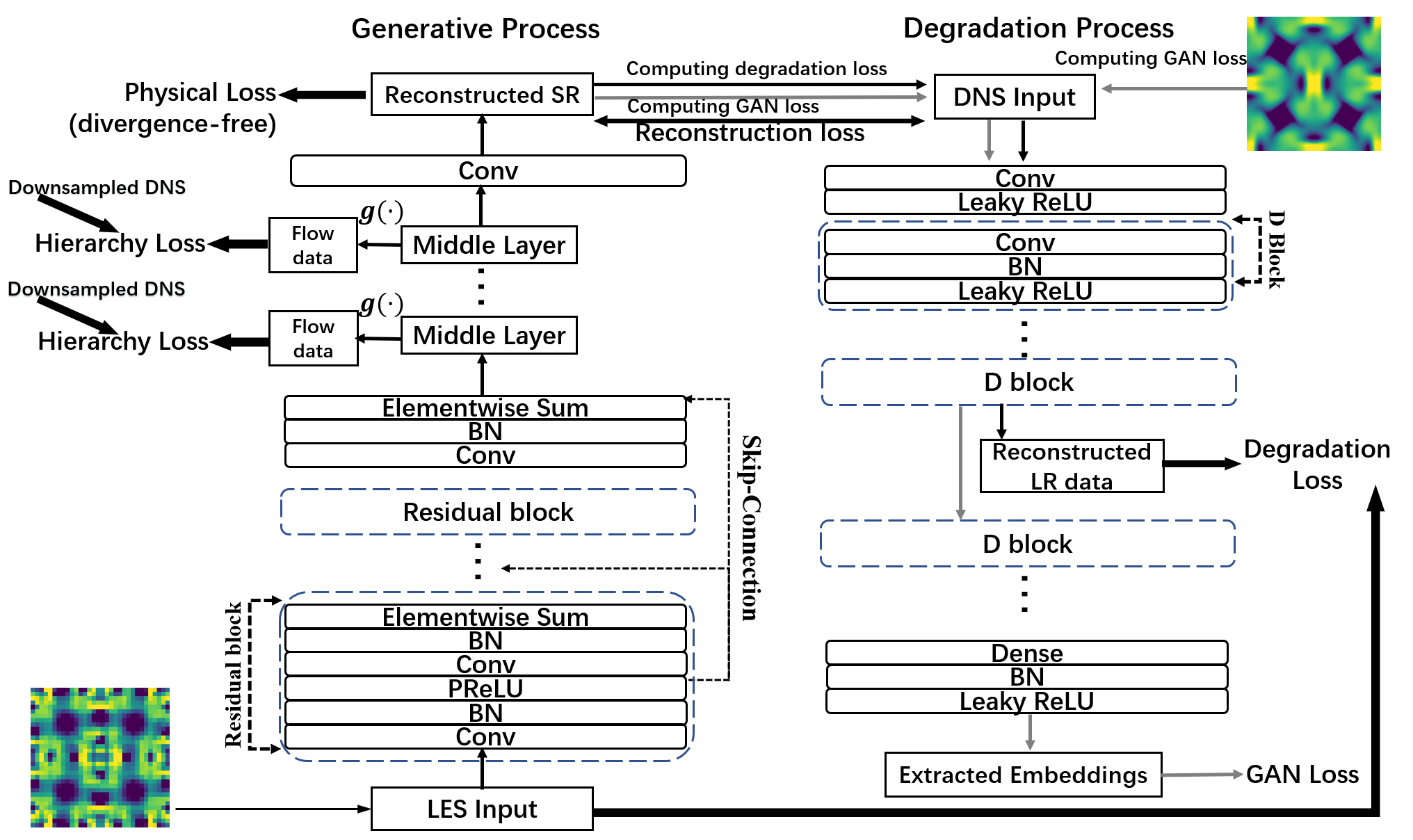}
\end{center}
\caption{The architecture of the proposed PGSRN model and different components in the loss function. Since the degradation structure is used for computing both the degradation loss and the GAN loss, we use the black arrows and grey arrows in the degradation process to show the flow for computing the degradation loss and GAN loss, respectively.}
\label{fig:tf_plot}
\end{figure*}

In this work, we develop a new method, termed Physics-Guided Super-Resolution Network (PGSRN)$\footnote{The source code for the PGSRN model presented in this study is available online at link: \url{https://drive.google.com/drive/folders/1w6j3pNzVqZ7Q9P7ZpnTsVNmvnfJXnxh_?usp=sharing}}$, to improve the reconstruction of high-resolution turbulent flow data. This development is by leveraging known physical constraints and explicitly exploring the discrepancy \& the consistency between different simulations. First, we generalize the loss function of the super-resolution model by incorporating the divergence-free velocity-field  constraint as required in incompressible flows.  Second, we introduce a hierarchical 
generative architecture by decomposing the data reconstruction into two steps:  (i)  transform low-resolution flow data into a down-sampled version of high-resolution data, and (ii) reconstruct high-resolution flow data from the down-sampled version. Step (i) allows explicitly modeling the data discrepancy due to different simulation methods used to generate low-resolution and high-resolution data. Step (ii) is to recover the fine-level details of flow data. Finally, we introduce a degradation process to further regularize reconstructed data by imposing the consistency between different simulations. Here we represent the degradation process by a forward model (by framing super-resolution as an inverse problem) that maps high-resolution data to low-resolution data. The forward model output of reconstructed data can then be compared against low-resolution simulations for consistency assessment. We further extend the degradation process as a feature extractor and introduce an adversarial loss on the extracted features from high-resolution data, which helps improve the modeling of fine-level fluid dynamics.  

For the purpose of demonstration, we consider a variant of the Taylor-Green vortex (TGV) \cite{brachet1984taylor}.  This is a three-dimensional incompressible flow and is simulated within a box with periodic boundary conditions. The TGV provides a suitable setting for our demonstration as it exhibits several salient features of turbulent transport. In this flow, the original vortex collapses into turbulent worm-like structures which become progressively more turbulent until viscosity eventually dissipates the large scale vortical structures. 
We compare our proposed method against several existing super-resolution algorithms to reconstruct DNS data of TGV. 
We also demonstrate the effectiveness of each component in our proposed method by showing the improvement both qualitatively and quantitatively. 

    

\section{Methodology}

The goal of our work is to achieve an end-to-end reconstruction mechanism from low-resolution LES data, denoted by $\textbf{X}_{LR}$ to high-resolution DNS data $\textbf{X}_{HR}$.  In Fig.~\ref{fig:tf_plot}, we show the overall structure of the methodology. The model has two components, the generative process and the degradation process. These are described here, in order: 

\subsection{Hierarchical Generative Process}

The generative process aims to map  $\textbf{X}_{LR}$ to $\textbf{X}_{HR}$. It contains multiple residual blocks and each block consists of convolutional layers~\cite{o2015introduction}, batch normalization layers~\cite{ioffe2015batch}, and parametric ReLUs following previous literature~\cite{he2015delving}. The generative process outputs a reconstructed data $\textbf{X}_{SR}$, and then the model is  optimized to reduce the difference between obtained $\textbf{X}_{SR}$ and provided high resolution data $\textbf{X}_{HR}$. Such a difference is represented as a reconstruction loss $\mathcal{L}_\text{recon}(\textbf{X}_{SR},\textbf{X}_{HR})$, which can be implemented as mean squared loss (MSE),  perceptual loss or other loss functions that measure the difference between two sets of data. In this work, we use the mean squared loss as we do not observe significant improvement using other loss functions.

\subsubsection{Hierarchical Structure}
We also build a hierarchical generative structure to decompose the information gap between low-resolution and high-resolution data and explicitly capture their difference. In particular, we consider two types of information loss from high-resolution data to low-resolution data: 1) the discrepancy caused by different simulation methods used to generate data of different scales, and 2) the loss of fine-level information due to the reduced resolution. 


Given the input low-resolution data $\textbf{X}_{LR}\in \mathbb{R}^{H\times W\times C}$ ($H$ and $W$ are spatial dimensions while $C$ is the number of physical variables), we use a hierarchical structure to extract intermediate data representation before generating high-resolution data of size ${KH\times KW\times C}$. In particular, we create multiple middle layers $\{\textbf{h}_1, \textbf{h}_2,. . . \textbf{h}_m\}$. Here the first middle layer $\textbf{h}_1$ is used to extract a down-sampled version of the high-resolution data $\textbf{X}_{HR}$, which is of size $H\times W$ (same with input low-resolution data).  By introducing this layer, the model can explicitly capture the discrepancy between simulation strategies used for generating the input data $\textbf{X}_{LR}$  (i.e., LES) and target data $\textbf{X}_{HR}$ (i.e., DNS) on the same resolution.  We introduce another loss on this middle layer to reduce the difference between extracted information from $\textbf{h}_1$ and the down-sampled $\textbf{X}_{HR}$.  
Specifically, the model first transforms the hidden layer $\textbf{h}_1$ into a reconstructed flow data of size $H\times W\times C$ via a function $g(\cdot)$ and then compare $g(\textbf{h}_1)$ against the down-sampled version of $\textbf{X}_{HR}$. 
More formally, this loss is expressed as:
\begin{equation}
\mathcal{L}_{h1} = \text{MSE}(g(\textbf{h}_1),\textbf{X}^{\text{down}_1}_{HR}),
\end{equation}
where $\textbf{X}^{\text{down}_1}_{HR}$ represents the down-sampled version of $\textbf{X}_{HR}$ of size $H\times W$.  
We implement the function $g(\cdot)$ using convolutional layers and fully connected layers. 

We can define additional losses on other intermediate layers $\textbf{h}_2,...,\textbf{h}_m$ by comparing down-sampled $\textbf{X}_{HR}$ using different down-sampling rates. Specifically, we represent the loss of the hierarchical generative process as follows:
\begin{equation}
\begin{aligned}
    \mathcal{L}_{hier} &= \alpha_1\mathcal{L}_\text{recon}(\textbf{X}_{SR},\textbf{X}_{HR})+ \alpha_2\sum_{i=1}^{m}\mathcal{L}_{hi} \\
    \mathcal{L}_{hi} &= \sum_{i=1}^m \text{MSE}(g(\textbf{h}_i),\textbf{X}^{\text{down}_i}_{HR})/m,
\end{aligned}
\end{equation}
where $\textbf{X}^{\text{down}_i}_{HR}$ is the down-sampled $\textbf{X}_{HR}$ of size $k_iH\times k_iW$, and $1=k_1<k_2<...<k_m<K$, and $\alpha_1$ and $\alpha_2$ are hyper-parameters to control the balance between the reconstruction loss on the predicted $\textbf{X}_{SR}$ and the loss on the middle layers. 

It is noteworthy that we only implement a two-dimensional super-resolution process in our tests, i.e., to increase the spatial dimensions from $H\times W$ to $KH\times KW$. For 3-D flow data, LES simulations often have coarser resolutions along two directions while keeping the other dimension to be the same. Hence, we can use the same method to reconstruct DNS simulations of size $KH\times KW\times D$ ($D$ for depths) using LES simulations of size $H\times W\times D$. 
The method presented herein can also be easily extended to include a three-dimensional convolutional filter if DNS simulations have higher resolution along all the three directions. 





\subsubsection{Physical Loss}
We further regularize the generative process by leveraging the physical constraints. 
These constraints can potentially reduce the size of the hypothesis space to be physically consistent, which helps extract more generalizable patterns and reduce the data required for training.


Specifically, we incorporate the inherent physical relationship. 
Here we represent the  velocity vector ${\bf V}({\bf x},t) $ along 3-D dimensions $({\bf x}\equiv x,y,z$),  by $u,\ v,$ and $w$, respectively. The flow is incompressible; thus, the  velocity field is divergence-free: 
\begin{equation}
\nabla \cdot {\bf V}= \frac{\partial u}{\partial x} + \frac{\partial v}{\partial y} + \frac{\partial w}{\partial z} = 0.
\end{equation}
Let ($u$, $v$, $w$) be included as three channels in $\textbf{X}_{LR}$ and $\textbf{X}_{HR}$. We use a second-order central finite difference approximation to estimate the partial derivatives.
We employ this divergent free property as an additional physical loss in the training process. 
\begin{equation}
\mathcal{L}_\text{Phy} = \sum_{(x,y,z)}\left[\nabla \cdot \hat{\bf V} ({\bf x},t)\right]^2/N,
\end{equation}
where $N$ is the number of spatial locations in the high-resolution data,  and $\hat{\bf V}$ represents the reconstructed velocity field at high resolution. 
By minimizing $\mathcal{L}_\text{Phy}$ on the velocity field, we penalize the reconstructed high-resolution flow data that significantly violate the divergence-free property. Such regularization can help reduce the search space for model parameters such that the reconstructed high-resolution data follow the divergence-free property which is enforced in incompressible flows. 




\subsection{Degradation Process} 


Given a low-resolution flow data sample, there can be multiple high-resolution samples that correspond to this low-resolution input. The generative process aims to find the mapping from low-resolution to high-resolution data that fit all the training samples, but are also prone to overfitting due to the large  high-resolution data space. On the other hand, given any high-resolution flow data samples, it would be much easier to learn a forward mapping which produces a unique correspondence at the coarser resolution. Here, we introduce a reverse degradation process to further regularize the model by considering the reconstruction as an inverse problem. Note that the degradation process cannot address the challenge of one-to-many correspondence from low-resolution to high-resolution data, but it can help eliminate reconstructed  
high-resolution flow data that are not consistent 
to the given low-resolution simulations. 
Such a degradation process is also helpful for capturing the discrepancy between simulation strategies used to generate data of different scales.


We create a forward model (in reverse modeling) $f(\cdot)$ that maps 
high-resolution $\textbf{X}_{HR}$ to low-resolution $\textbf{X}_{LR}$.  We implement this forward model by using stacked convolutional layers and batch normalization layers (see Fig.~\ref{fig:tf_plot}).  Then we introduce an additional degradation loss to ensure the consistency between the given low-resolution data $\textbf{X}_{LR}$, and the low resolution data obtained from the reconstructed $\textbf{X}_{SR}$ through the forward model, i.e., $f(\textbf{X}_{SR};\theta_\text{deg})$, where $\theta_\text{deg}$ represents model parameters in the forward model.  In particular, we implement the degradation loss as follows:

\begin{equation}
    \mathcal{L}_\text{deg} = \text{MSE}(\textbf{X}_{LR},f(\textbf{X}_{SR};\theta_\text{deg}))
\end{equation}



The parameters of the forward model in the degradation process are estimated by minimizing the degradation loss. Moreover, we introduce the additional GAN loss~\cite{goodfellow2014generative} by sharing the architecture of the forward model and the discriminator used in the GAN-based model. In particular, the GAN loss is defined on the extracted features by further extending the forward model, as shown in Fig.~\ref{fig:tf_plot}. The GAN-based loss has been shown to improve the performance of extracting high-resolution textures in super-resolution tasks~\cite{Cheo2003SR}.

Formally, the loss function of the degradation process is:
\begin{equation}
\mathcal{L}_{D} = \beta_1 \mathcal{L}_\text{deg} + \beta_2 \mathcal{L}_\text{GAN,disc}, 
\end{equation}
where $\mathcal{L}_\text{GAN,disc}$ is the discriminator loss used in SRGAN~\cite{ledig2017photo}, $\beta_1$ and $\beta_2$ are hyper-parameters.

The generative process needs to be optimized in conjunction with the degradation process. In particular, it is optimized by minimizing the combination of reconstruction loss (including middle layers) in the hierarchical structure, the physical Loss, degradation loss, and the GAN loss (the generator part). The overall loss for the generative process is:
\begin{equation}
\begin{aligned}
\mathcal{L}_{G} &= \alpha_1\mathcal{L}_\text{recon}(\textbf{X}_{SR},\textbf{X}_{HR})+ \alpha_2\sum_{i=1}^{m+1}\mathcal{L}_{hi} + \alpha_3 \mathcal{L}_\text{Phy}\\
&+\alpha_4 \mathcal{L}_\text{deg}+\alpha_5 \mathcal{L}_\text{GAN,gen},
\end{aligned}
\end{equation}
where 
$\mathcal{L}_\text{GAN,gen}$ is the standard generator loss in
GAN~\cite{ledig2017photo}. We have also explored other extensions of GAN-based loss functions, such as Wasserstein GAN~\cite{arjovsky2017wasserstein} and Wasserstein GAN-GP~\cite{gulrajani2017improved} but did not observe significant improvement. The hyper-parameters \{$\alpha_{1:5}$\} are used to control the weight of each component. We discuss the selection of these parameters in Section~\ref{sec:exp}.




\section{Experiment}
\label{sec:exp}

In this section, we first introduce the dataset and  experimental settings. Then we evaluate the performance of our proposed method in reconstructing DNS data. 

\subsection{Experiment Setting}
We consider two experiments: single-snapshot  and cross-time. The former is designed to verify the ability of the proposed method in reconstructing a new data portion over space , and the latter is for evaluation of the methodology for data reconstruction over time. 

\subsubsection{Dataset}

The TGV is produced by solution of the constant density Navier-Stokes equation:

\begin{equation}
    \frac{\partial \textbf{V}}{\partial t} + (\textbf{V}. \nabla) \textbf{V} = \frac{-1}{\rho} \nabla p + \nu \Delta \textbf{V},
\end{equation}
where $\rho ({\bf x},t)$ and $p ({\bf x},t)$ denote the fluid density and the thermodynamic pressure, respectively. 
The evolution of the TGV includes enhancement of vorticity stretching and the consequent production of small-scale eddies. Initially, large vortices are placed in a cubic periodic domain of $[-\pi,\pi]$ (in all three-directions), with initial conditions: 
\begin{eqnarray}
u (x,y,z,0) &=& \sin(x) \cos(y) \cos(z) \\ v(x,y,z,t) &=& - \cos(x)\sin(y)\cos(z) \\ w(x,y,z,t) &=& 0.    
\end{eqnarray}
Then the value of the Reynolds number is set to $Re=1600$.  
We have LES and DNS results of TGV at several times steps.  
For each time step, we consider the three-components of the velocity along the $x$, $y$ and $z$ axis, denoted by  $u$, $v$ and $w$, respectively.
Our objective is to reconstruct the DNS results of 
the velocity field $(u,v,w)$ using LES data.  In particular, $\textbf{X}_{LR}$ represents the LES predicted values of the velocity field while the target $\textbf{X}_{HR}$ represents the DNS results of the velocity field. 
Here both LES and DNS data are generated along 65 grid points along the $z$ axis under equal intervals. The LES and DNS  are conducted on 32-by-32 and 128-by-128 grid points, respectively, 
along the $xy$ 
directions. Hence, the DNS data is of 4 times higher resolution compared to LES data.  

\subsubsection{Evaluation Metrics}
We evaluate the performance of DNS reconstruction  using two different metrics, root mean squared error (RMSE) and structural similarity index measure (SSIM)~\cite{wang2004image}. We use RMSE to measure the difference (error) between reconstructed data and target DNS data. The lower value of RMSE indicates better reconstruction performance. 
SSIM  is used to appraise the similarity between reconstructed data and target DNS on three aspects, luminance, contrast and overall structure. 

\subsubsection{Baselines}
We compare the performance of PGSRN method against several existing  methods that have been widely used for image super-resolution and turbulent flow downscaling. Specifically, we implement SRCNN~\cite{dong2014learning}, RCAN~\cite{zhang2018image}, SRGAN~\cite{ledig2017photo}, and a popular dynamic fluid downscaling method: DCS/MS~\cite{fukami2019super} as baselines. 

To better verify the effectiveness of each component in our proposed method, we further compare PGSRN with two of its variants: PGSRN-P and PGSRN-H as described below. 

\textit{The variant with only physical Loss (PGSRN-P): } To show the effectiveness of the physical loss, we remove the degradation Loss and hierarchical loss (in middle layers) from the Hierarchical Generative Process. We name this method as  PGSRN-P.

\textit{The variant with physical loss + hierarchical generative process (PGSRN-H):} In this baseline, we  remove only the degradation loss from the PGSRN method, and we name this baseline as PGSRN-H.

By comparing PGSRN-P and SRGAN, we hope to show the improvement by incorporating the physical loss. We can further verify the effectiveness of the hierarchical loss by comparing PGSRN-P and PGSRN-H. Finally, the comparison between PGSRN-H and the complete version of PGSRN can show the effectiveness of using the degradation loss. 

\begin{figure*} [!h]
\centering
\subfigure[LES. ]{ \label{fig:a}
\includegraphics[width=0.3\columnwidth]{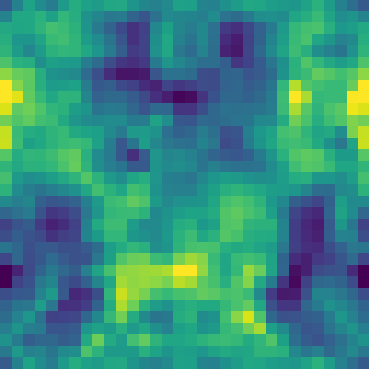}
}\hspace{5mm}
\subfigure[Upsampling.$\backslash$ 0.505]{ \label{fig:b}
\includegraphics[width=0.3\columnwidth]{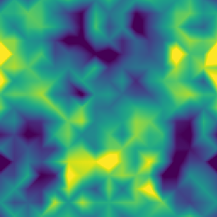}
}\hspace{5mm}
\subfigure[DCS/MS.$\backslash$ 0.640]{ \label{fig:b}
\includegraphics[width=0.3\columnwidth]{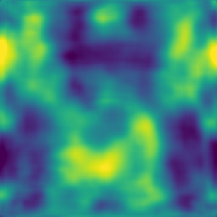}
}\hspace{5mm}
\subfigure[SRCNN.$\backslash$ 0.643]{ \label{fig:b}
\includegraphics[width=0.3\columnwidth]{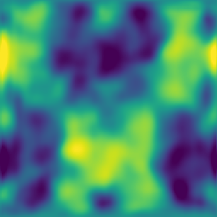}
}\hspace{5mm}
\subfigure[RCAN.$\backslash$ 0.565]{ \label{fig:b}
\includegraphics[width=0.3\columnwidth]{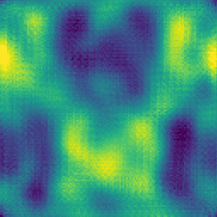}
}

\subfigure[SRGAN.$\backslash$ 0.659]{ \label{fig:b}
\includegraphics[width=0.3\columnwidth]{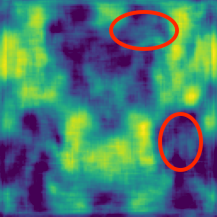}
}\hspace{5mm}
\subfigure[PGSRN-P.$\backslash$ 0.710]{ \label{fig:b}
\includegraphics[width=0.3\columnwidth]{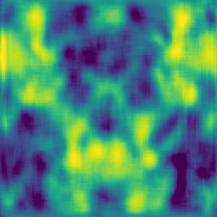}
}\hspace{5mm}
\subfigure[PGSRN-H.$\backslash$ 0.712]{ \label{fig:b}
\includegraphics[width=0.3\columnwidth]{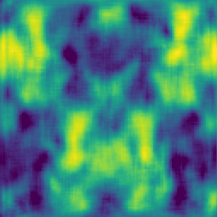}
}\hspace{5mm}
\subfigure[PGSRN.$\backslash$ 0.762]{ \label{fig:b}
\includegraphics[width=0.3\columnwidth]{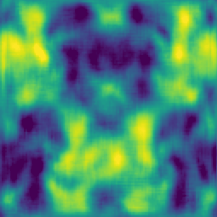}
}\hspace{5mm}
\subfigure[Target DNS.]{ \label{fig:b}

\includegraphics[width=0.3\columnwidth]{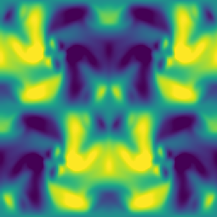}
}
\vspace{.08in}
\caption{Reconstructed $u$ channel by each method on a sample testing slice along the $z$ dimension in the single-snapshot experiment. We also show the SSIM value for each reconstructed data.}
\vspace{.1in}
\label{fig:tf_plot1}
\end{figure*}

\begin{table}[!t]
\small
\newcommand{\tabincell}[2]{\begin{tabular}{@{}#1@{}}#2\end{tabular}}
\centering
\caption{Reconstruction performance (measured by RMSE and SSIM) on $(u,v,w)$ channels by different methods in the single-snapshot experiment.}
\begin{tabular}{l|cccc}
\hline
\textbf{Method} & RMSE & SSIM &  \\ \hline 
SRCNN & (0.086, 0.089, 0.106) &(0.833, 0.833, 0.764)&\\ 
RCAN & (0.096, 0.095, 0.109) &(0.745, 0.741, 0.645)&\\  
DSC/MS & (0.096, 0.095, 0.106) &(0.828, 0.826, 0.729)&\\ 
SRGAN & (0.089, 0.078, 0.085) &(0.837, 0.834, 0.751)&\\    
\hline
PGSRN-P & (0.076, 0.075, 0.075)   & (0.845, 0.846, 0.781)&\\ 
PGSRN-H & (0.077, 0.074, 0.070) &(0.844, 0.844, 0.800)&\\ 
PGSRN & (0.064, 0.061, 0.066) &(0.875, 0.877, 0.838)&\\ 
\hline
\end{tabular}
\label{fig:table1}
\end{table}

\subsubsection{Experimental Design}

We evaluate the performance of our proposed method in two different scenarios. First, we consider the case in which part of flow data is missing at a specific time. For example, the high-resolution flow data is available at certain points along the $z$ axis but not available at other points. We can use the model trained using available data to reconstruct high-resolution DNS data for the remaining locations. 
We refer to this test as a single-snapshot experiment since the training and testing are conducted at the same time step. 
In this test, we use the 5-fold cross validation method to divide 65 data slices into five parts and each part has 13 slices. Each time we use four folds (i.e., 52 slides) for training and use the remaining one fold (i.e., 13 slides) for testing. 

Second, we conduct cross-time experiments to study how the proposed method helps simulate flows in a dynamic scenario. 
In particular, we use data from  20 consecutive time steps  as the training data, and then test the model in the next 20 time steps. 
This is a challenging task since dynamic fluid is changing over time following complex non-linear patterns (driven by Navier-Stokes equation~\cite{foias2001navier}). Hence, the model trained from available data may not be able to generalize to future data that look very different with training data. 

\subsubsection{Training Settings}
Data normalization is performed on both training and testing datasets, to normalize input LES variables to the range [0,1]. Then, the model is trained by ADAM optimizer~\cite{kingma2014adam_arxiv}  with $\beta_1$ = 0.5, $\beta_2$ = 0.999. The initial learning rate is set to 0.0002 and iterations are 500 epochs. 
All the $\alpha_{i}$ ($i=1$ to 5) values are set to 1. We use Tensorflow 1.15 and Keras to implement our models with Titan Xp GPU.

\begin{table}[!t]
\small
\newcommand{\tabincell}[2]{\begin{tabular}{@{}#1@{}}#2\end{tabular}}
\centering
\caption{Reconstruction performance on $(u,v,w)$ channels by RMSE and SSIM in the cross-time experiment. 
The performance is measured at a time step which is 5 seconds apart from the last time step in training data.
}
\begin{tabular}{l|cccc}
\hline
\textbf{Method} & RMSE & SSIM &  \\ \hline 
SRCNN&(0.089, 0.089, 0.122) &(0.890, 0.889, 0.848)&\\ 
RCAN& (0.073, 0.073, 0.093)&(0.875, 0.874, 0.837)&\\ 
DSC/MS&(0.095, 0.098, 0.131) &0.881, 0.879, 0.822)&\\
SRGAN& (0.086, 0.087, 0.096) &(0.897, 0.901, 0.860)&\\  

\hline
PGSRN-P&(0.086, 0.082, 0.095)&(0.903, 0.909, 0.876)&\\ 
PGSRN-H&(0.083, 0.081, 0.093) &(0.907, 0.908, 0.870)&\\
PGSRN&(0.076, 0.072, 0.086) &(0.920, 0.922, 0.896)&\\
\hline
\end{tabular}
\label{fig:table2}
\end{table}

\subsection{Single-Snapshot Experiment}

\textbf{Quantitative Results.}
Table.~\ref{fig:table1} shows quantitative comparisons amongst all the methods. 
When comparing our proposed PGSRN method with baseline methods, our proposed PGSRN performs the best on both evaluation ways, obtaining lowest RMSE value and highest SSIM values. According to this table, the proposed method PGSRN in general outperforms other baselines for velocity components \{$u$, $v$, $w$\} in terms of both RMSE and SSIM. 

By comparing SRGAN  and  SRCNN, we show the improvement by using the GAN loss. Furthermore, the comparison amongst SRGAN, PGSRN-P, PGSRN-H, and PGSRN shows the effectiveness of incorporating each component (physical loss, hierarchical loss, degradation) of the proposed method. In particular, the incorporation of physical loss and degradation loss brings the most significant performance improvement in terms of RMSE and SSIM.

Although the baseline DSC/MS was developed in the context of turbulent flow downscaling, it has been  successful only when tested towards reconstructing DNS data using down-sampled DNS data. This method does not work well in our test because it does not take into account the discrepancy between LES and DNS results.

\textbf{Visual Results.} In Fig.~\ref{fig:tf_plot1}, we show an example of high-resolution flow data (128-by-128 on a specified $z$ value) reconstructed by each method. We observe that LES results on this slice do not capture some fine-level details of DNS. The DCS/MS, SRCNN, and RCAN methods also do not capture such fine-level information. In contrast, SRGAN can obtain higher SSIM value compared to these methods, and it can be further improved by incorporating physical loss, hierarchical loss, and the degradation loss. By visually inspecting reconstructed data over multiple slices, we find that the incorporation of the physical in general helps eliminate artifacts that are physically inconsistent (e.g., the red circled areas in Fig.~\ref{fig:tf_plot1} (f)). The use of degradation loss generally helps better match the magnitude of the reconstructed data to the target DNS. 
For most slices for which LES exhibits obvious differences with DNS data, the SSIM value of PGSRN is more than 10$\%$ higher than all the other existing algorithms.

\subsection{Cross-Time Experiment}

\textbf{Quantitative Results.} In this experiment, we compare the same set of existing methods as in the previous single-snapshot experiment. In Table~\ref{fig:table2}, we show the quantitative results at the 5th time step in the testing set, which is five seconds apart from the last time step in the training set. We select this time step to show the performance in short-term prediction while we will show more results at different time steps later in the temporal analysis.

We can observe similar results that our proposed PGSRN outperforms other methods by a considerable margin. We also notice that RCAN method achieves good  RMSE in this test, e.g., RCAN has smaller RMSE than PGSRN in reconstructing velocity components $u$ and $v$. This shows that the machine learning model with a more complex (and carefully designed) forward process has a better chance at matching with the target data. However, the success of RCAN is limited in capturing the overall structure of flows as observed from its lower SSIM values compared with other methods.

\begin{figure*} [!h]
\centering
\subfigure[$u$ Channel.]{ \label{fig:a}{}
\includegraphics[width=0.32\linewidth]{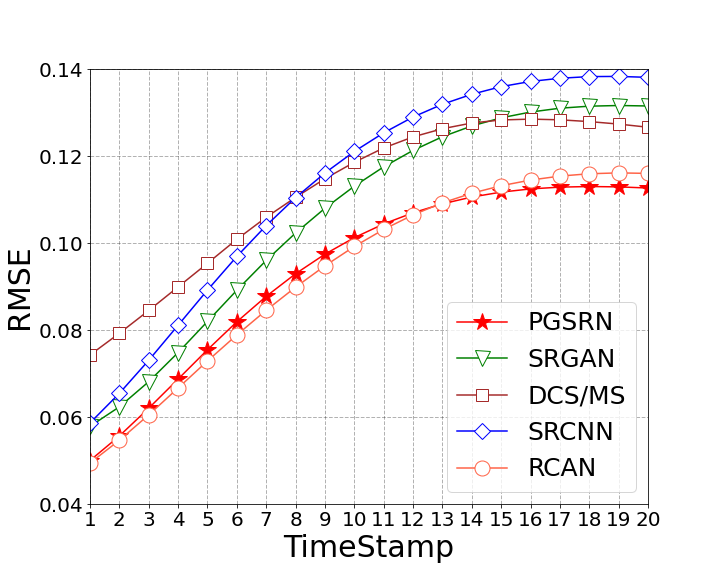}
}\hspace{-0.2in}
\subfigure[$v$ Channel.]{ \label{fig:b}{}
\includegraphics[width=0.32\linewidth]{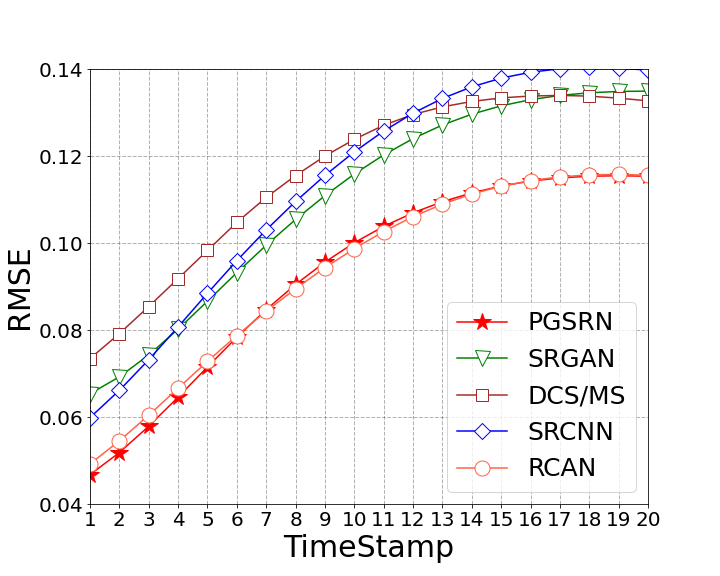}
}\hspace{-0.2in}
\subfigure[$w$ Channel.]{ \label{fig:b}{}
\includegraphics[width=0.32\linewidth]{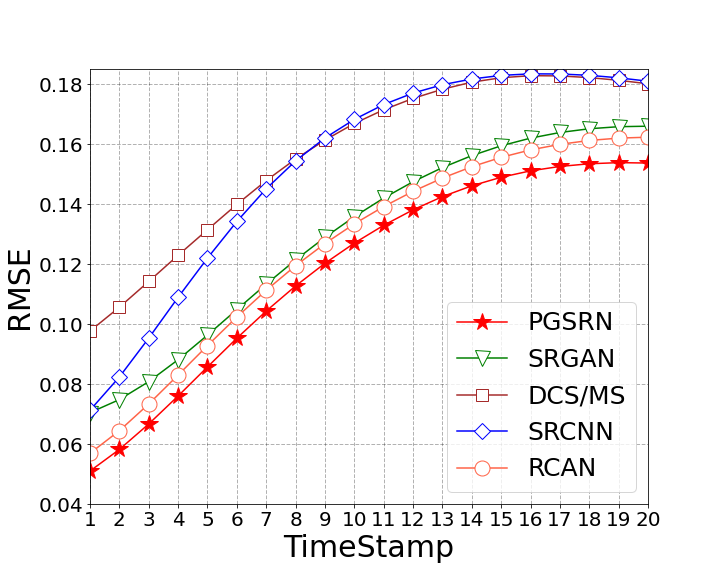}
}
\vspace{.05in}
\caption{Change of RMSE values produced by different models from the 1st to 20th time steps in the cross-time experiment.}
\label{fig:tf_plot3}
\end{figure*}

\begin{figure*} [!h]
\centering
\subfigure[$u$ Channel.]{ \label{fig:a}{}
\includegraphics[width=0.32\linewidth]{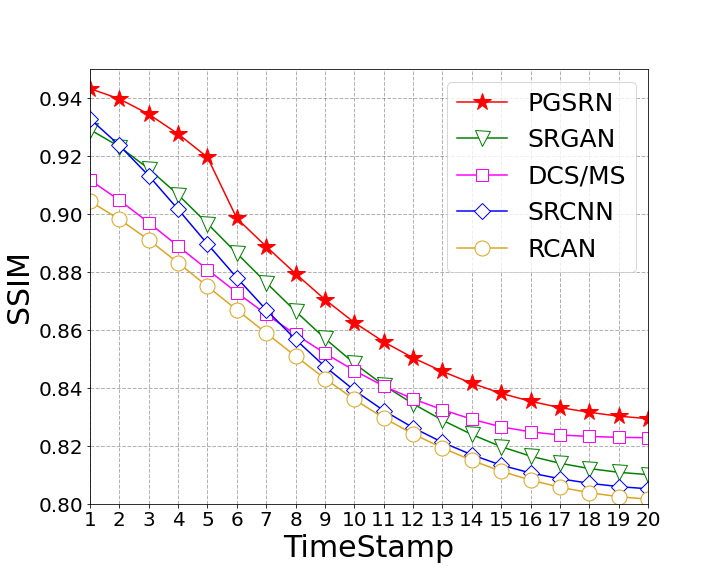}
}\hspace{-0.2in}
\subfigure[$v$ Channel.]{ \label{fig:b}{}
\includegraphics[width=0.32\linewidth]{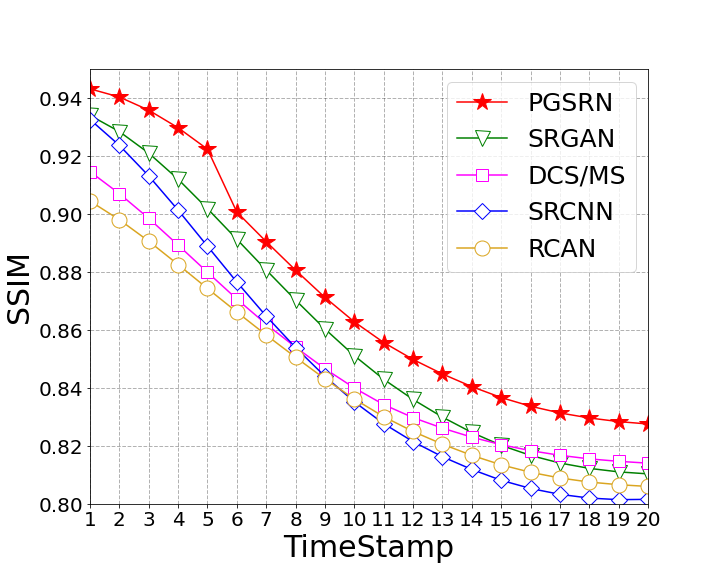}
}\hspace{-0.2in}
\subfigure[$w$ Channel.]{ \label{fig:b}{}
\includegraphics[width=0.32\linewidth]{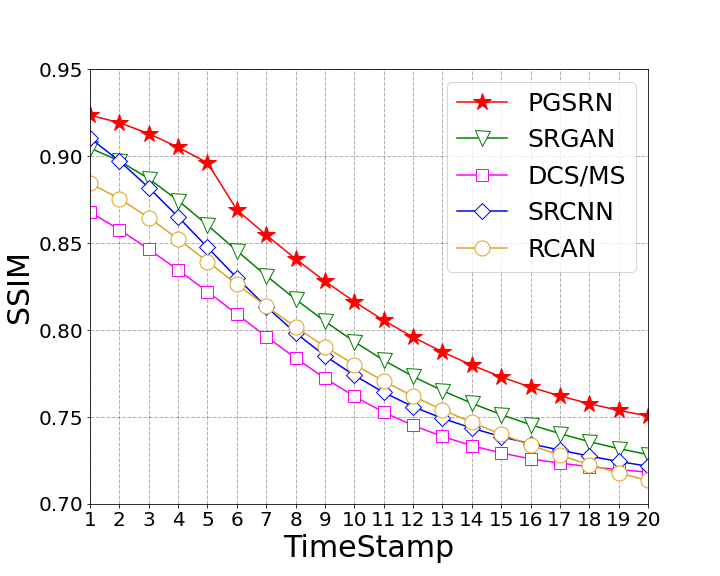}
}
\vspace{.05in}
\caption{Change of SSIM values produced by different models from the 1st to 20th time steps in the cross-time experiment.}
\label{fig:tf_plot4}
\end{figure*}

\textbf{Temporal Analysis.} 
In the temporal analysis, we show the change of performance as we reconstruct DNS data over 20 time steps after the training data.  
We show the performance of cross-time prediction in terms of RMSE and SSIM  in Fig.~\ref{fig:tf_plot3} and Fig.~\ref{fig:tf_plot4}, respectively. We have several observations from these figures: (1) With larger time intervals between training data and prediction data, the performance (in terms of both RMSE and SSIM) becomes worse. In general, our method still has better performance than other methods.  (2) RCAN has achieved better RMSE than the proposed method for reconstructing $u$ and $v$ channels, especially during the first 10 time steps. However, our method has much better SSIM compared to RCAN.   (3) After 15 time steps, the performance (SSIM/RMSE) tends to be stable;   (4)  Although our proposed PGSRN method achieves better RMSE and SSIM than other methods, the reconstructed data is of very low quality when the time gap is large. We will show some examples in the visual results.


\begin{figure*} [!h]
\centering
\subfigure[DCS/MS.$\backslash$ 0.749]{ \label{fig:a}
\includegraphics[width=0.15\linewidth]{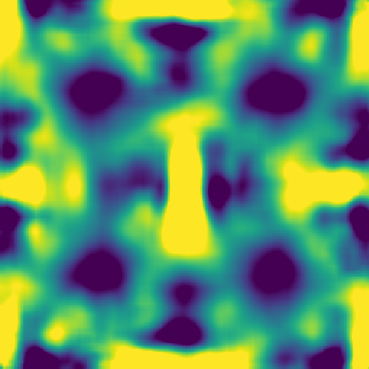}
}\hspace{5mm}
\subfigure[SRCNN.$\backslash$ 0.839]{ \label{fig:b}
\includegraphics[width=0.15\linewidth]{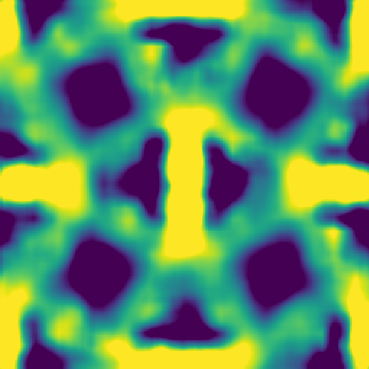}
}\hspace{5mm}
\subfigure[SRGAN.$\backslash$ 0.858]{ \label{fig:c}
\includegraphics[width=0.15\linewidth]{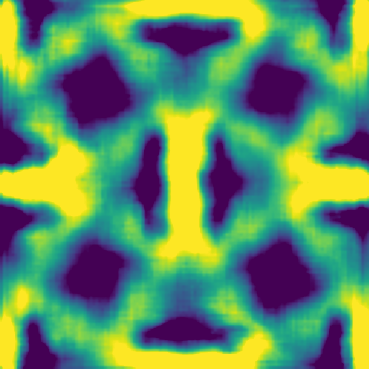}
}\hspace{5mm}
\subfigure[PGSRN.$\backslash$ 0.886]{ \label{fig:d}
\includegraphics[width=0.15\linewidth]{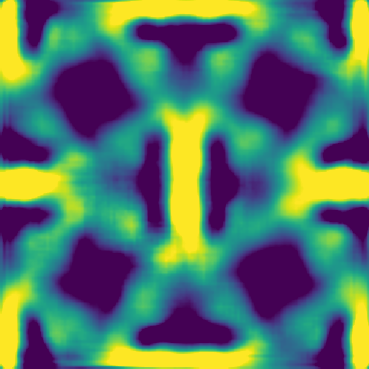}
}\hspace{5mm}
\subfigure[Target DNS.]{ \label{fig:e}
\includegraphics[width=0.15\linewidth]{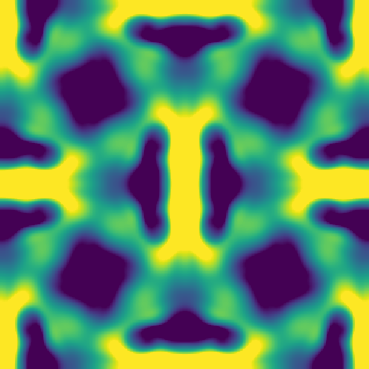}
}\hspace{5mm}
\subfigure[DCS/MS.$\backslash$ 0.715]{ \label{fig:a}
\includegraphics[width=0.15\linewidth]{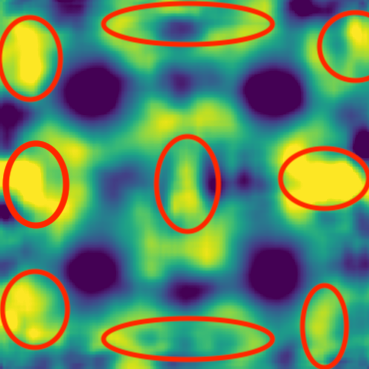}
}\hspace{5mm}
\subfigure[SRCNN.$\backslash$ 0.724]{ \label{fig:b}
\includegraphics[width=0.15\linewidth]{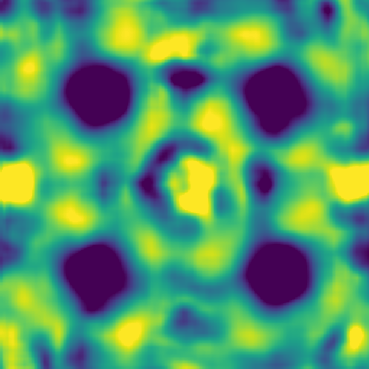}
}\hspace{5mm}
\subfigure[SRGAN.$\backslash$ 0.802]{ \label{fig:b}
\includegraphics[width=0.15\linewidth]{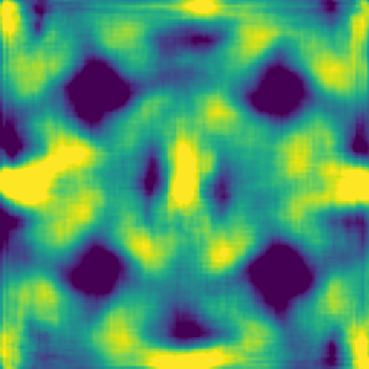}
}\hspace{5mm}
\subfigure[PGSRN.$\backslash$ 0.857]{ \label{fig:b}
\includegraphics[width=0.15\linewidth]{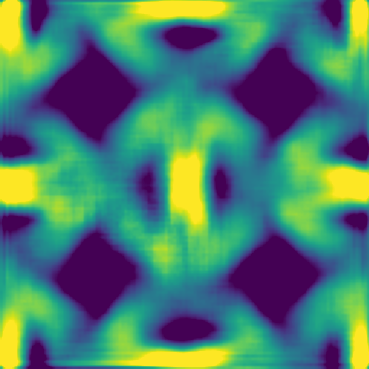}
}\hspace{5mm}
\subfigure[Target DNS.]{ \label{fig:b}
\includegraphics[width=0.15\linewidth]{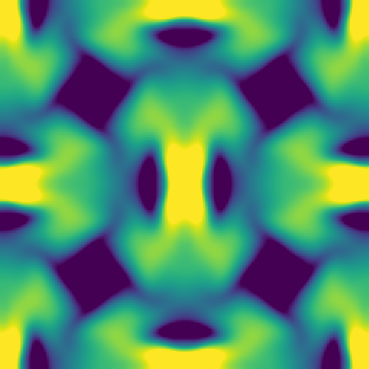}
}\hspace{5mm}
\subfigure[DCS/MS.$\backslash$ 0.597]{ \label{fig:a}
\includegraphics[width=0.15\linewidth]{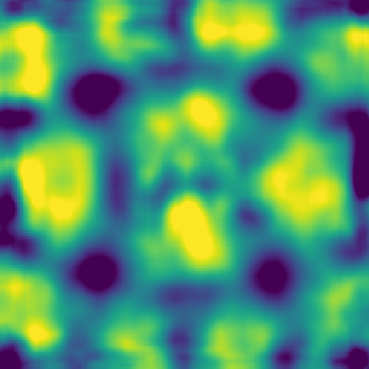}
}\hspace{5mm}
\subfigure[SRCNN.$\backslash$ 0.642]{ \label{fig:b}
\includegraphics[width=0.15\linewidth]{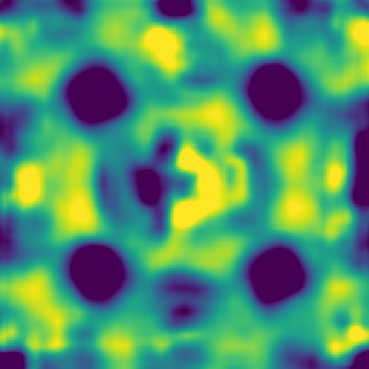}
}\hspace{5mm}
\subfigure[SRGAN.$\backslash$ 0.658]{ \label{fig:b}
\includegraphics[width=0.15\linewidth]{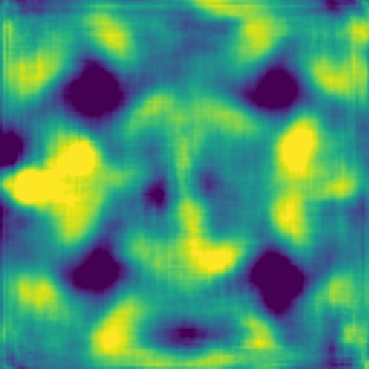}
}\hspace{5mm}
\subfigure[PGSRN.$\backslash$ 0.688]{ \label{fig:b}
\includegraphics[width=0.15\linewidth]{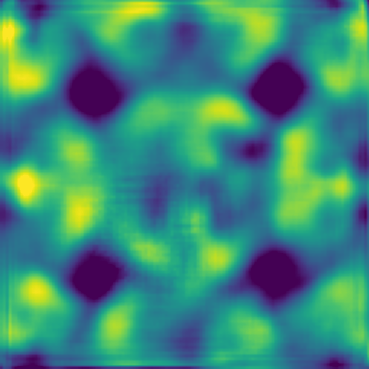}
}\hspace{5mm}
\subfigure[Target DNS.]{ \label{fig:b}
\includegraphics[width=0.15\linewidth]{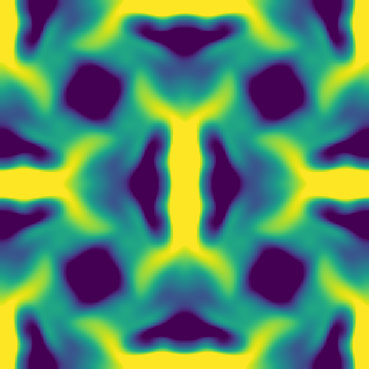}
}\hspace{5mm}
\subfigure[DCS/MS.$\backslash$ 0.656]{ \label{fig:a}
\includegraphics[width=0.15\linewidth]{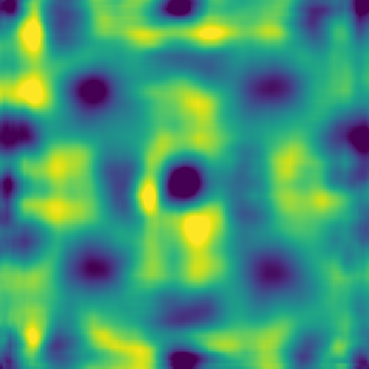}
}\hspace{5mm}
\subfigure[SRCNN.$\backslash$ 0.670]{ \label{fig:b}
\includegraphics[width=0.15\linewidth]{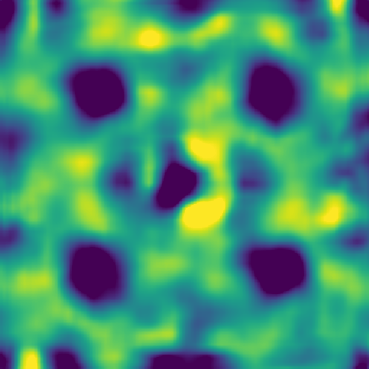}
}\hspace{5mm}
\subfigure[SRGAN.$\backslash$ 0.728]{ \label{fig:b}
\includegraphics[width=0.15\linewidth]{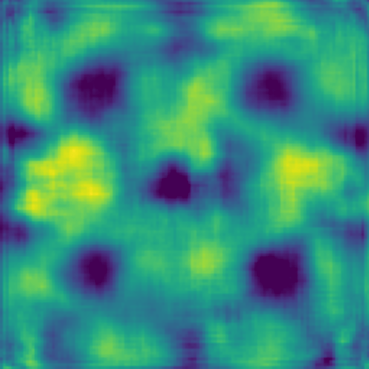}
}\hspace{5mm}
\subfigure[PGSRN.$\backslash$ 0.756]{ \label{fig:b}
\includegraphics[width=0.15\linewidth]{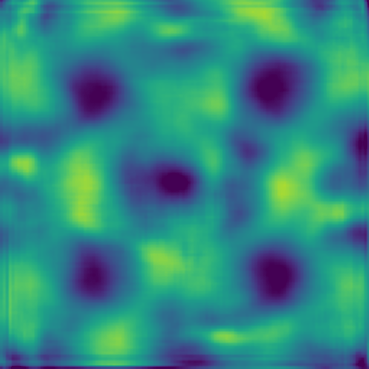}
}\hspace{5mm}
\subfigure[Target DNS.]{ \label{fig:b}
\includegraphics[width=0.15\linewidth]{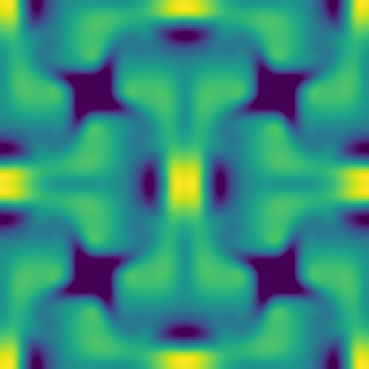}
}
\vspace{.05in}
\caption{Reconstructed $w$ channel by each method on a sample testing slice along the $z$ dimension in the cross-time experiment. We show the reconstruction results at the 1st time step, 5th time step, 10th time step and 20th time step in (a)-(e), (f)-(j), (k)-(o) and (p)-(t), respectively.  We also show the SSIM value for each reconstructed data.}
\label{fig:tf_plot2}
\end{figure*}

\textbf{Visual Results.} In Fig.~\ref{fig:tf_plot2}, we show the reconstructed data at multiple time steps (1st time step, 5th time step,  10th time step and 20th time step) after the training period. For each time step, we only show the slice of the $w$ component at a specified $z$ value. 
At the 1st step, both our method and other methods can obtain ideal reconstruction results. This is because the test data is similar to the training data at the last time step. According to reported SSIM values, our proposed method is slightly better than other baseline methods. At the 5th time step, 
our proposed PGSRN method performs much better than other methods. This confirms that our method can effectively eliminate the differences between input picture and ground truth picture, 
accurately fine-level capture textures and patterns (e.g., see red circled areas), reduce color amplitude difference, and thus achieve much better performance. At the 10th and 20th time steps, since the testing data is very different from training data, 
neither our method nor other methods can provide good reconstruction results. 

One potential limitation of our proposed method and other methods is that they mainly focus on the reconstruction using the spatial information, and pay less attention to  temporal dependencies. Hence, these models may not fully capture fluid dynamics transport over time. After a sufficiently long time gap, the dynamic flow data can become very different from the data used in model training. 
Hence, it is difficult for either our proposed methods or other state-of-the-art methods to obtain a relative positive consequence. We will keep this as our future work to further preserve long-term consistency to underlying fluid dynamics (e.g., by following the Navier-Stokes equation).

\textbf{Validation based on Physical Metrics.} We also show the performance of cross-time prediction in terms of Reynolds stress~\cite{Nawab2020}, which is considered an important metric for studying the property of turbulence. In particular, Reynolds stress is computed as:  
\begin{equation}
 \mathcal{R}_{ab} = \overline{ab} - \bar{a}\bar{b}   
\end{equation}
where $a$ and $b$ represent any velocity variables 
from $(u, v, w)$, and $\bar{a}$ represents the mean value of variable $a$ over the entire field. 

\begin{figure*} [!t]
\centering
\subfigure[$\mathcal{R}_{uv}$.]{ \label{fig:a}{}
\includegraphics[width=0.32\linewidth]{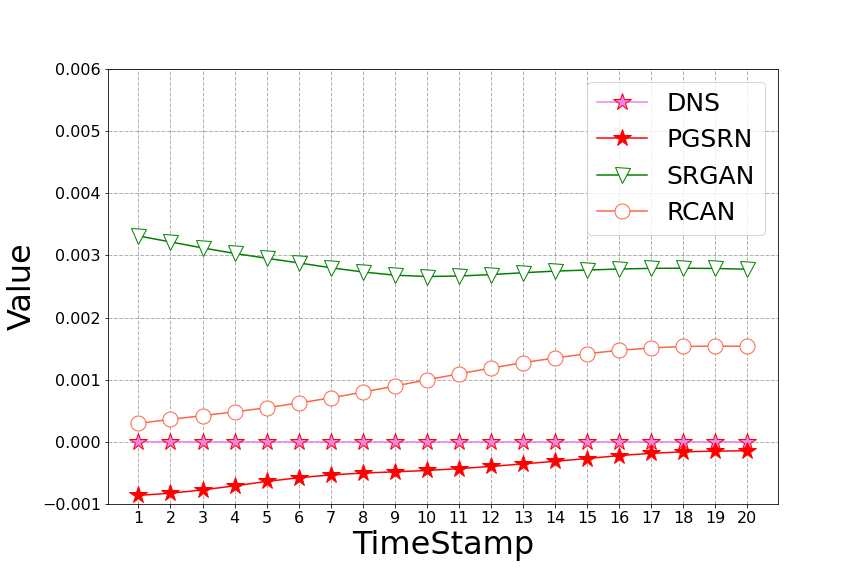}
}\hspace{-0.2in}
\subfigure[$\mathcal{R}_{uw}$]{ \label{fig:b}{}
\includegraphics[width=0.32\linewidth]{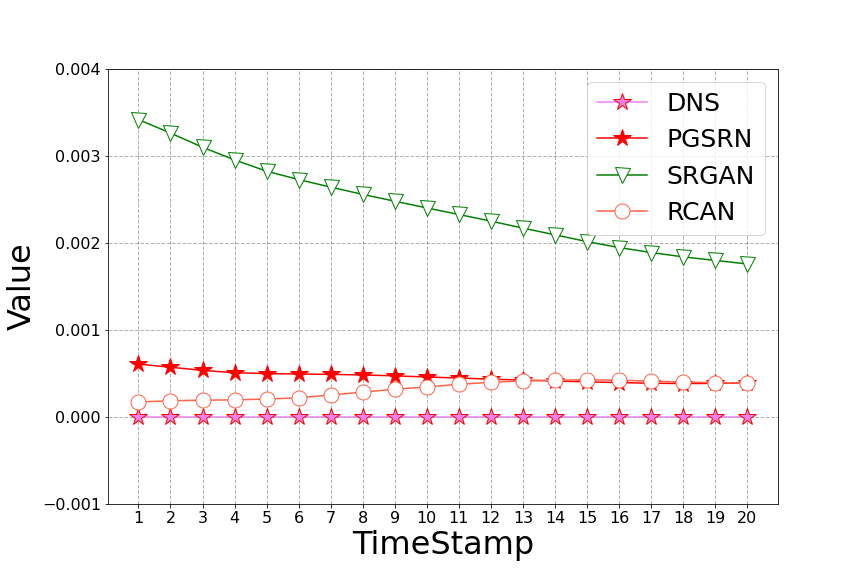}
}\hspace{-0.2in}
\subfigure[$\mathcal{R}_{vw}$.]{ \label{fig:b}{}
\includegraphics[width=0.32\linewidth]{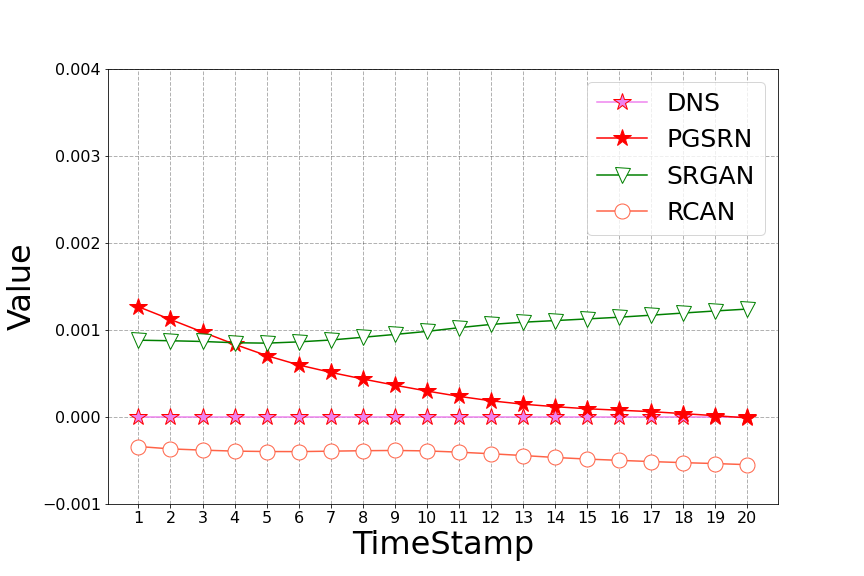}
}
\vspace{.05in}
\caption{Change of Reynolds Stress values produced by the reference DNS and different models from the 1st to 20th time steps in the cross-time experiment.}
\label{fig:tf_plot6}
\end{figure*}

In Fig.~\ref{fig:tf_plot6}, we show the Reynolds stress of target DNS and Reynolds stress of reconstructed flow data by RCAN, SRGAN, and our method over time. Ideally, high-quality reconstruction should have Reynolds stress similar to that of the DNS data. However, this may not be true in practice since 
different flow data can have the similar Reynolds stress values and the super-resolution model does not directly optimize the similarity of Reynolds stress during the training process. 
In this figure, we observe that SRGAN performs poorly since its Reynolds stress values are far away from those of the DNS data. Our method PGSRN performs similarly to RCAN on estimating the three components of Reynolds stresses. 
However, there is still a large discrepancy between our method and the real DNS data in terms of Reynolds stresses. Sophisticated machine learning models are prone to produce artificial factors in reconstructed flow resulting in unreliable Reynolds stress. In the future, we will pursue optimizing the super-resolution model by including physical metrics in the training objective.  

\subsection{Parameter Sensitivity}
\begin{figure*} [!t]
\centering
\subfigure[The value of $\alpha_3$.]{ \label{fig:a}{}
\includegraphics[width=0.51\linewidth]{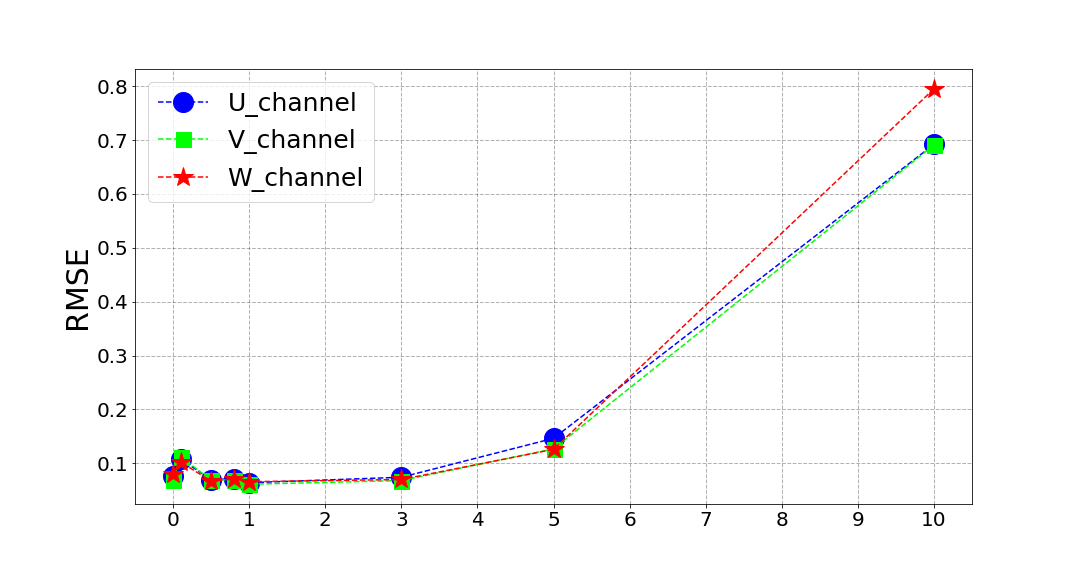}
}\hspace{-0.5in}
\subfigure[The value of $\alpha_4$.]{ \label{fig:b}{}
\includegraphics[width=0.51\linewidth]{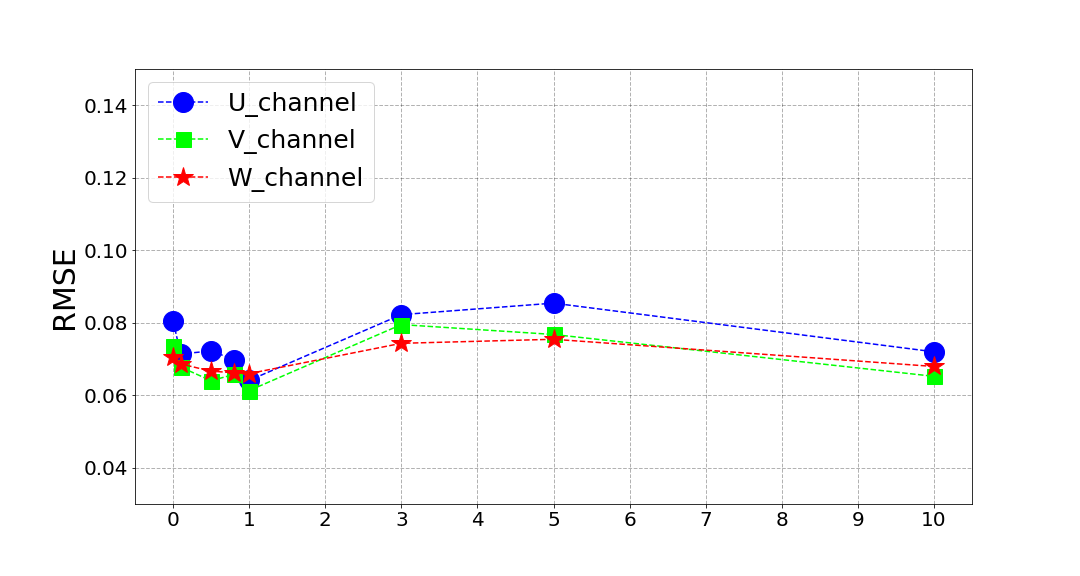}
}

\caption{Change of RMSE as we adjust the hyper-parameters in the loss function. (a) The variation of performance (RMSE) with different values of  $\alpha_{3}$, i.e., the weight for the physical loss. (b) The variation of performance (RMSE) with different values of  $\alpha_{4}$, i.e., the weight for the degradation loss.}
\label{fig:tf_plot5}
\end{figure*}

We also test the performance of the proposed method using different hyper-parameters in the loss function. In particular, we report the RMSE achieved by our method in reconstructing the three velocity components $\{u,v,w\}$ when varying the weights of physical loss ($\alpha_{3}$) and degradation loss($\alpha_{4}$), 
as shown in Fig.~\ref{fig:tf_plot5}. When we change the value of one hyper-parameter, we keep all the other $\alpha_i$ values as 1. 

For the value of $\alpha_3$, when it increases from 0 to 1, the model gets better performance due to the contribution of physical regularization. However, as we keep increasing the value of $\alpha_3$, especially when it is greater than 3, the model starts to produce worse performance. In the physical loss, we use the finite difference approximation for the divergence and thus the simulations may not strictly follow this regularization. When we set a larger $\alpha_3$ value, the training process is dominated by the physical loss while paying less attention to other loss terms, which leads to a degraded reconstruction performance.

For the weight of the degradation loss, we can observe similar patterns that the model performs better when $\alpha_4$ increases from 0 to 1. Unlike the physical loss, when we further increase the value of $\alpha_4$, the performance is relatively stable. The performance becomes worse when we increase $\alpha_4$ from 1 to 5, but becomes better from 5 to 10.  


\section{Related Work}
In this section, we introduce related literature on several topics. We start with existing super-resolution methods that are widely used in computer vision, which is followed by their adaptations to the problem of turbulent flow simulations. Finally, we  discuss existing works on incorporating physical relationships into the loss function of machine learning models.

\subsection{Machine Learning for Super-resolution in Computer Vision}


Researchers have developed many deep learning-based methods for single image super resolution (SISR) in computer vision. The neural network structures, such as convolutional network layers, are known to be able to extract spatial contextual information that is needed for recovering  high-resolution data. The recent advances in GAN-based methods also enables better extracting high-resolution textures that are similar to target data.  

One of the earliest models that uses deep convolutional networks for SISR problem is SRCNN~\cite{dong2014learning}. SRCNN can directly learn the end-to-end mapping between coarse-resolution and high-resolution images using a series of convolutional layers. 
Compared to SRCNN, Residual Channel Attention Network (RCAN)~\cite{zhang2018image} uses a very deep trainable structure with additional skip-connection layers.  The intuition of RCAN is to  bypass the abundant low-frequency information and focus more on the relevant information. The skip-connections are also known to improve the stability of the optimization process for deep neural networks. Moreover, RCAN
rescales features of each channel to fully explore  the interdependencies among channels. 
Recently, there are also other popular methods based on residual structures such as HDRN~\cite{Duong2021}, SAN~\cite{Dai2019}, RDN~\cite{zhang2018residual}, CARN~\cite{ahn2018fast}, and DRRN~\cite{Tai2017}.

Another popular super-resolution model is the SRGAN model~\cite{ledig2017photo}, which employs generative adversarial network (GAN) for the SISR problem.  SRGAN model not only stacks the deep residual network to build a deeper generative network for image super resolution, but also introduces 
a discriminator network to distinguish reconstructed images and real images using an adversarial loss function. The ultimate goal is to train the generative network such that the reconstructed images cannot be easily distinguished by the discriminator. 
Compared with other models, one major advantage of the SRGAN model is that the discriminator can help extract representative features from high-resolution data and enforce such features in the reconstructed images. 
Recently, there are also other extensions to the SRGAN method that further improve the performance~\cite{chen2017fsrnet,wang2018recovering, wang2018esrgan,karras2018progressive,gan8759375,cheng2021mfagan,Long2021}.

These super-resolution methods have shown success in benchmark image datasets, but they are not designed for capturing complex patterns amongst multiple physical variables and the discrepancy between different simulation methods. Hence, they may lead to unsatisfactory performance in reconstructing flow data, especially when the resolution ratio between input data and target data is large.

\subsection{Machine Learning for Reconstructing Flow Data} Given the importance of simulating high-resolution flows, there is a surge of interest in using super-resolution techniques for reconstructing high-resolution flow data. 
Fukami et al.~\cite{fukami2019super} propose an improved CNN-based hybrid DSC/MS model by extracting patterns from multiple scales. This method has been shown to produce good performance on reconstructing the turbulent velocity and vorticity fields from extremely low-resolution input data. This model has also shown success to handle spatio-temporal super resolution analysis in turbulent flow~\cite{Fukami_2020}.

Similarly, Liu et al.~\cite{liu2020deep} also propose another CNN-based model MTPC to simultaneously handle spatial and temporal information in turbulent flow simultaneously to fully capture features in different time ranges. 
There are also other approaches that are inspired by GAN. For example, Xie et al. \cite{xie2018tempogan} introduce tempoGAN, which augments a general adversarial network with an additional discriminator network along with additional loss function terms that preserve temporal coherence in the generation of physics-based simulations of fluid flow.  Deng et al.~\cite{Deng2019SuperresolutionRO} demonstrate that both SRGAN and ESRGAN~\cite{wang2018esrgan} can 
produce good reconstruction of  high-resolution turbulent flow in their datasets.

Most of these existing methods on reconstructing flow data still rely on simple CNN-based structure and do not leveraging recent advances in the super-resolution. Hence, they are limited in their capacity to extract complex non-linear relationships. Moreover, most of these approaches have only shown success in data reconstruction using a down-sampled version of the target data. These methods do not take into account the discrepancy of different simulations (e.g., LES and DNS) and thus may have degraded performance in our problem. 

\subsection{Physics-based Loss Function}

When applied to scientific problems, standard machine learning models can fail to capture complex relationships amongst physical variables, especially when provided with limited observation data. This is one reason for their failure to generalize to scenarios not encountered in training data. Hence, researchers are beginning to incorporate physical knowledge into loss functions to help machine learning models capture generalizable dynamic patterns consistent with established physical relationships.


The use of physical-based loss functions have already shown promising results in a variety of scientific disciplines. In a recent survey~\cite{willard2020integrating}, Willard et al. summarize existing literature and approaches for incorporating scientific knowledge into machine learning models. For example, in lake water temperature modeling, Karpatne et al.~\cite{karpatne2017physics} propose an additional physics-based penalty based on known monotonic physical  relationship to guarantee that the density of water at lower depth is always greater than the density of water in any depths above. 
Then, Jia et al.~\cite{jia2019sdm2} and Read et al.~\cite{read2019process} further extended this work by including an additional penalty term on violating the law of energy conservation. In the problem related to vortex induced vibrations, Kahana et al.~\cite{Adar2020physical} apply an additional loss function to ensure the physical consistency in the time evolution of waves, which has been used in an inverse problem about distinguishing an underwater obstacle's location from acoustic measurements. This additional physical-based loss function has been shown to improve the prediction results and makes the model more robust. 

Another application is to solve the PDEs of dynamical systems. Researchers commonly use physical-based loss functions for the mandatory compliance with the governing equation in the loss function. Raissi et al.~\cite{raissi2019physics} develop data-efficient spatial-temporal function approximators to solve PDEs and estimate PDE parameters.  
Similarly,an encoder-decoder is proposed by Zhu et al.~\cite{Zhu2019physical} to predict transient PDE by controlling PDE constraints. 


\section{Discussion and Future works}
In this paper, we develop a new data-driven method PGSRN that leverages physical relationships to fully explore the reconstruction gaps between coarse-resolution and high-resolution simulations of fluid dynamics. Specifically, we leverage the underlying physical relationships to regularize the relationships amongst velocity components in flow data. To further explore the correspondence and discrepancy between DNS and LES data, we also  build the hierarchical generative process and introduce the degradation process. 
We have demonstrated the effectiveness of our proposed method in reconstructing DNS of flow data from coarse-resolution LES data 
through both single-snapshot  and cross-time experiments. Compared with existing methods, the proposed method can better recover fine-level fluid patterns that are missing from coarse-resolution LES data, and thus produce better performance in both tests. We have also shown that all the components introduced in our proposed method are helpful in the reconstruction process.



Although our method has been developed in the context of simulating fluid dynamics, the involved techniques can be widely used for other important scientific problems. For example,  simulations of cloud-resolving models (CRM) at sub-kilometer horizontal resolution are critical for effectively representing boundary-layer eddies and low clouds. However, it is not feasible to generate simulations at such fine resolution even with the most powerful commuters expected to be available in the near future. Hence, the method developed in this paper can provide a great potential for reconstructing high-resolution simulations.

Additionally, as shown in the cross-time experiment, our method remains limited in reconstructing long-term data. This requires new mechanisms to enforce underlying physical processes on fluid dynamics (e.g., Navier-Stokes equation). Furthermore, we plan to introduce other related parameters besides velocity (e.g., mass and pressure) to supplement and further optimize our model. Last, we also will introduce other domain's metrics (e.g., Reynolds Stress and Kinetic Energy~\cite{Nawab2020}) as the training loss to enhance the trustworthiness of the model when it is deployed for long-term and large-scale simulations. We will purse these directions in our future work.


\bibliographystyle{IEEEtran}
\bibliography{IEEEabrv-SC}



\end{document}